\documentclass[prl,twocolumn,superscriptaddress]{revtex4-1}

\usepackage{graphicx}
\usepackage{color}
\usepackage{amssymb}
\usepackage{bm}
\usepackage{amsmath,amsfonts,latexsym}

\usepackage{braket} 
\usepackage{enumitem}
\usepackage{comment}

\usepackage{bm} 

\newcommand*{\cE}{{\cal E}}

\newcommand*{\cT}{{\cal T}}

\newcommand*{\cV}{{\cal V}}

\newcommand*{\bop}{b^{\vphantom{\dagger}}}
\newcommand*{\bdop}{b^\dagger}

\begin{document}

\title{From GPE to KPZ: finite temperature dynamical structure factor of the 1D Bose gas}
\author{Manas Kulkarni}
\affiliation{Department of Physics, University of Toronto, Ontario, M5S 1A7, Canada}
\affiliation{Department of Physics, University of Virginia,
Charlottesville, VA 22904-4714 USA}
\author{Austen Lamacraft} 
\affiliation{Department of Physics, University of Virginia,
Charlottesville, VA 22904-4714 USA}
\date{\today}
\date{\today}

\begin{abstract}
We study the finite temperature dynamical structure factor $S(k,\omega)$ of a 1D Bose gas using numerical simulations of the Gross--Pitaevskii equation appropriate to a weakly interacting system. The lineshape of the phonon peaks in $S(k,\omega)$ has a width $\propto |k|^{3/2}$ at low wavevectors. This anomalous width arises from resonant three-phonon interactions, and reveals a remarkable connection to the Kardar--Parisi--Zhang universality class of dynamical critical phenomena.
\end{abstract}

\maketitle

The statistical mechanics of low dimensional fluids, both quantum and classical, has long been a source of theoretical surprises. To give just two examples:
\begin{enumerate}[itemsep=2pt,parsep=1pt]
	\item The long-time tail $\propto t^{-d/2}$ in the velocity autocorrelation function of a $d$-dimensional classical fluid invalidates hydrodynamics for $d\leq 2$ \cite{Alder:1970,Ernst:1970,Dorfman:1970}.
	
	\item The Luttinger liquid description \cite{Haldane:1981} provides a universal language for 1D quantum liquids, with a panoply of phases arising upon perturbation \cite{Giamarchi:2004}. 
\end{enumerate}

Despite these achievements recent developments in the theory of 1D quantum liquids away from the low energy limit make it clear that our understanding of these systems is still rather limited (for a review, see Ref.~\cite{Imambekov:2011}). To take a simple example, consider the dynamical structure factor $S(k,\omega)$ that gives the cross section for inelastic scattering from the liquid as a function of momentum $\hbar k$ and energy $\hbar\omega$ transferred. The Luttinger liquid theory predicts that $S(k,\omega)$ consists only of a pair of delta function peaks $\omega = \pm c|k|$, with $c$ the velocity of sound, corresponding to an undamped phonon oscillation.  At finite $k$, however, one expects this delta function to broaden due to interactions between phonons. Attempts to find the resulting lineshape using perturbation theory within the Luttinger framework are plagued by divergences \cite{Aristov:2007}, whose origin we will describe below. As a result, the possibility of capturing the relevant physics within the Luttinger or hydrodynamic formalism is now viewed with a degree of pessimism \cite{Cheianov:2009}.

\begin{figure}
	\centering
	\includegraphics[width=\columnwidth]{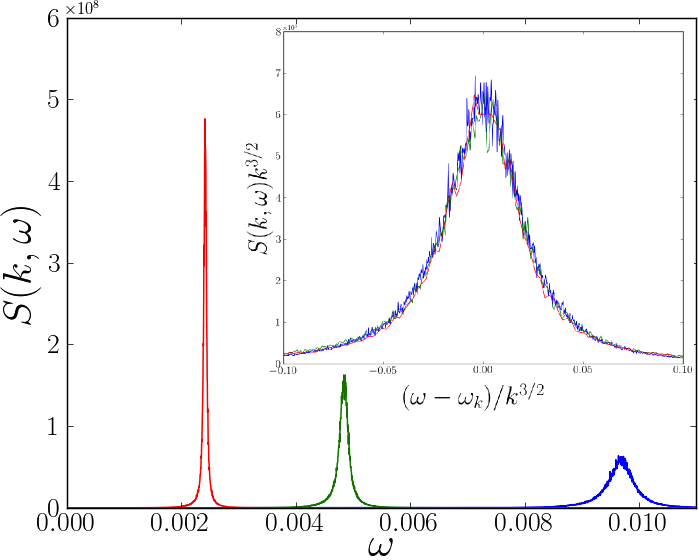}
	\caption{Dynamical structure factor $S(k,\omega)$ of a 1D Bose gas described by the Gross--Pitaevskii equation for wavevectors $k=2\pi p/L$, $p=64,32,16$ (right to left). $L=5\times 2^{13}$ and temperature $T=0.005$, with length being measured in units of the healing length, and energy in units of the chemical potential. Inset: Scaling collapse of the phonon peaks using the ansatz Eq.~\eqref{GPEtoKPZ_PhononScaling}.}
	\label{fig:Lineshape}
\end{figure}

Almost all of the developments reviewed in Ref.~\cite{Imambekov:2011} pertain to zero temperature. In this work we study the dynamical structure factor of a 1D Bose gas at finite temperature. Aside from being of paramount importance in real systems, we will show that finite temperature brings qualitatively new features that cannot be interpreted simply as a smearing of the zero temperature lineshape. Using analytical arguments and simulations of the Gross--Pitaevskii equation (GPE) appropriate to weak interactions and finite temperature $T$, we find that the lineshape of the phonon peak in $S(k,\omega)$ has a width $\Gamma_{k}\propto |k|^{3/2}$ at low $k$ (see Fig.~\ref{fig:Lineshape}). As well as dominating any zero temperature structure (generally $\propto k^{2}$) at low wavevectors, the $3/2$ power is \emph{anomalous} relative to the $k^{2}$ scaling that follows from linearized hydrodynamics \cite{Forster:1975}.

This unusual scaling points to a very rich phenomenology. According to a remarkable recent conjecture \cite{beijeren2011}, the long wavelength dynamics of a classical 1D fluid at finite temperature is in the Kardar--Parisi--Zhang (KPZ) universality class describing interface growth \cite{Kardar:1986,kriecherbauer2010,sasamoto2010}. Specifically, the phonon (Brillouin) peaks in $S(k,\omega)$ have the scaling form at low wavenumber
\begin{equation}
	\label{GPEtoKPZ_PhononScaling}
	S^{(\pm)}_{\text{phonon}}(k,\omega) \propto \frac{1}{\Gamma_{k}}f_{\text{PS}}\left(\frac{\omega\pm c|k|}{\Gamma_{k}}\right)
\end{equation}
where $f_{\text{PS}}(x)$ is given in Eq.~(5.7) of Ref.~\cite{prahofer2004}. The meaning of Eq.~\eqref{GPEtoKPZ_PhononScaling} is that in a frame moving at the speed of sound the density fluctuations moving in the same direction behave exactly as the fluctuations of the interface slope in the KPZ problem. 

There are very few experiments confirming KPZ scaling to date \cite{wakita1997,maunuksela1997,takeuchi2010}. Our hope is that the results of this Letter will lead to its observation in new systems. For example, the structure factor of 1D Bose gases of $^{87}$Rb was recently measured using Bragg spectroscopy \cite{Fabbri:2011}, while in Ref.~\cite{savard2011} the hydrodynamics of superfluid Helium in a single nanohole was investigated. In the latter case the sound absorption coefficient is presumably more accessible than the structure factor, and being $\propto\Gamma_{k}$ displays the same anomalous scaling.


%

\emph{Hydrodynamic description}. Our starting point is the \emph{classical} Hamiltonian describing a 1D gas of bosons of mass $m$ and interaction parameter $g$
\begin{equation}
	\label{GPEtoKPZ_GPHam}
	H = \int dx\,\left[\frac{|\partial_{x}\Psi|^{2}}{2m}+\frac{g}{2}|\Psi|^{4}\right],
\end{equation}
where the complex field $\Psi(x)$ obeys the Poisson bracket $\left\{\Psi^{\dagger}(x),\Psi(y)\right\}=i\delta(x-y)$, and we have set $\hbar=1$. The dynamics of $\Psi(x,t)$ is described by the familiar Gross--Pitaevskii equation. The mean field description embodied by the GPE is appropriate when the number of particles in a healing length $\xi\equiv\left(g\rho_{0}m\right)^{-1/2}$ is large, where $\rho_{0}$ denotes the mean density. This corresponds to `Luttinger parameter' $K\equiv\frac{\pi \rho_{0}}{mc}\gg 1$,  with $c=\sqrt{g\rho_{0}/m}$ the speed of sound in the uniform state. 

After writing the condensate order parameter as $\Psi(x)=\sqrt{\rho(x)}e^{i\theta(x)}$ in terms of the canonically conjugate density $\rho(x)$ and phase $\theta(x)$, the Hamiltonian takes the form
\begin{equation}
	\label{GPEtoKPZ_HydroHam}
	H = \int dx \left[\frac{\rho\left(\partial_{x} \theta\right)^{2}}{2m}+\frac{(\partial_{x} \sqrt{\rho})^{2}}{2m}+\frac{g}{2}\rho^{2}\right].
\end{equation}
Dynamics near a state of uniform density with $\rho(x,t)=\rho_{0}$, $\theta(x,t)=0$ can be described in the first approximation by writing $\rho=\rho_{0}+\varrho$, retaining only terms quadratic in $\varrho$ and $\theta$ from Eq.~\eqref{GPEtoKPZ_HydroHam}
\begin{equation}
	\label{GPEtoKPZ_QuadHam}
	H_2 = \int dx \left[\frac{\rho_{0}\left(\partial_{x} \theta\right)^{2}}{2m}+\frac{(\partial_{x} \varrho)^{2}}{8m\rho_{0}}+\frac{g}{2}\varrho^{2}\right].	
\end{equation}
$H_{2}$ is solved by introducing the mode expansions for $\varrho(x)$ and $\theta(x)$ for a system of length $L$
\begin{equation}
	\label{Chris_notes_genmodes}
	\begin{split}
\varrho(x)&=\sqrt{\frac{\rho_{0}}{2L}}\sum_{k\neq 0}e^{-\kappa_{k}}\left(b_{k}e^{ikx}+\text{c.c}\right)	\\
	\theta(x)&= \frac{i}{\sqrt{2\rho_{0}L}} \sum_{k\neq 0}e^{\kappa_{k}}\left(b_{k}e^{ikx}-\text{c.c}\right).
	\end{split}
\end{equation}
After substitution in Eq.~\eqref{GPEtoKPZ_QuadHam}, $e^{\kappa_{k}}$ is chosen to diagonalize $H_{2}=\sum_{k} \cE_{k}|\bop_{k}|^{2}$
%
%
%
with $\cE_{k}=\left[\frac{k^{2}}{2m}\left(\frac{k^{2}}{2m}+2g \rho_{0} \right)\right]^{1/2}$ the Bogoliubov dispersion relation. At low $k$ $\cE_{k}\to c|k|+O(k^{3})$. The deviation from the linear dispersion is due to the second term of Eqs.~\eqref{GPEtoKPZ_HydroHam} and \eqref{GPEtoKPZ_QuadHam}, sometimes called the `quantum pressure'.

Interactions between the modes are described by the anharmonic parts of Eq.~\eqref{GPEtoKPZ_HydroHam}. The most important interaction arises from the first term, and has the form
\begin{equation}
	\label{GPEtoKPZ_CubicVertex}
	\begin{split}
		H_{3} &= \int dx\, \frac{\varrho(\partial_{x}\theta)^{2}}{2m}\\
	&=\sum_{k_{1}+k_{2}+k_{3}=0}\sqrt{\frac{c|k_{1}k_{2}k_{3}|}{32L\rho_{0}m}}\left(\bop_{k_{1}}\bop_{k_{2}}\bop_{k_{3}}\right.\\
	&\qquad\qquad\left.-\bop_{k_{1}}\bdop_{-k_{2}}\bdop_{-k_{3}}-\bdop_{k_{1}}\bop_{-k_{2}}\bop_{-k_{3}}\right).
	\end{split}
\end{equation}
where for simplicity we have assumed the low $k$ limit for $e^{\kappa_{k}}$. The difficulty associated with a perturbative treatment of this interaction is now apparent. Substituting the time dependence $\bop_{k}\to\bop_{k}e^{-i\cE_{k}t}$ associated with $H_{2}$, we see that the second and third terms of Eq.~\eqref{GPEtoKPZ_CubicVertex} are \emph{resonant} for purely linear dispersion when all three modes move in the same direction (in quantum mechanical language energy and momentum conservation are simultaneously satisfied). One may object that the $O(k^{3})$ deviation from linearity at low $k$ due to the quantum pressure term removes this difficulty, but the $|k|^{3/2}$ broadening that we find dominates this effect at low $k$. The other nonlinearities arising from Eq.~\eqref{GPEtoKPZ_HydroHam} are likewise irrelevant in this limit.

The need for a non-perturbative approach was recognized long ago in Ref.~\cite{andreev:1980}, where a self-consistent mode-coupling (SCMC) treatment of the cubic interaction was given, ignoring vertex corrections, and yielded $\Gamma_{k}\propto \sqrt{T |k|^{3}}$, where $T$ is the temperature (see also Ref.~\cite{Samokhin:1998}). The same result was independently rederived much later \cite{delfini2006}. In contrast, a renormalization group (RG) argument based on Galilean invariance (first appearing in the related context of the noisy Burgers equation \cite{forster1977}) predicts a dynamical critical exponent $z=1+d/2$ for $d<2$ \cite{narayan:2002}. This suggests that Galilean invariance lies behind the success of the SCMC approach, an idea that finds support in the analysis of vertex corrections for the case of Burgers equation \cite{frey1996}.

We wish to emphasize that the SCMC theory of Refs.~\cite{andreev:1980,Samokhin:1998,delfini2006} is an uncontrolled approximation, while the RG analysis of Ref.~\cite{narayan:2002} was based on the equations of viscous 1D hydrodynamics with thermal fluctuations appearing as noise sources. It is therefore desirable to study the purely Hamiltonian dynamics described by Eq.~\eqref{GPEtoKPZ_GPHam}.

\begin{figure}
	\centering
		\includegraphics[width=0.8\columnwidth]{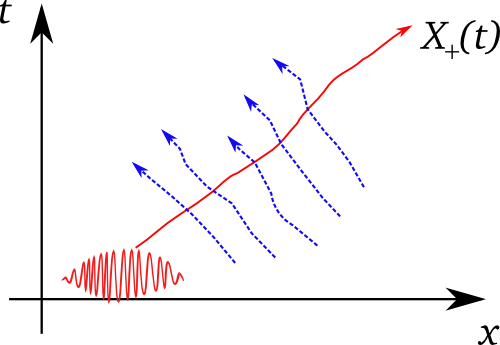}
	\caption{The characteristic curve $X_{+}(t)$ giving the path of a right-moving phonon wavepacket is not straight due to the influence of the counterpropoagating waves}
	\label{fig:characteristics}
\end{figure}

\emph{Equations of motion.} To gain some intuition regarding the connection of the GPE to the noisy Burgers equation and thence to the KPZ universality class, we discuss the equations of motion of the Hamiltonian Eq.~\eqref{GPEtoKPZ_HydroHam}. Ignoring the quantum pressure term, these are
\begin{equation}
	\label{GPEtoKPZ_HydroEq}
	\begin{split}
	\partial_{t} \rho+\partial_{x}\left(v\rho\right)=0	\\
	\partial_{t}v + v\partial_{x}v+(g/m)\partial_{x}\rho=0
	\end{split}
\end{equation}
where $v=\partial_{x}\theta/m$ is the superfluid velocity. These are the continuity and Euler equations for the 1D Bose gas, and may be put in the Riemann form \cite{Menikoff:1989}.
\begin{equation}
	\label{GPEtoKPZ_RHproblem}
	\partial_{t}(v\pm 2c_{\rho})+v_{\pm}\partial_{x}(v\pm 2c_{\rho})=0,
\end{equation}
where $v_{\pm}=v\pm c_{\rho}$, and $c_{\rho}=c\sqrt{\rho/\rho_{0}}$ is the speed of sound in a frame in which the fluid is locally at rest. Eq.~\eqref{GPEtoKPZ_RHproblem} tells us that $v\pm 2c_{\rho}$ are constant along their respective characteristic curves $X_{\pm}(t)$ defined by $\dot X_{\pm}(t)=v_{\pm}(X_{\pm}(t),t)$. Alternatively, we may write Eq.~\eqref{GPEtoKPZ_RHproblem} as
\begin{equation}
	\label{GPEtoKPZ_2Burgers}
	\partial_{t}v_{\pm}+v_{\pm}\partial_{x}v_{\pm}=\frac{1}{3}(\partial_{t}+v_{\pm}\partial_{x})v_{\mp}.
\end{equation}
The interpretation of Eq.~\eqref{GPEtoKPZ_2Burgers} is as follows. Right and left moving sound waves propagate through the fluid at the velocities $v_{\pm}$, but their motion is perturbed by the variation in the comoving frame of the velocity of the counterpropagating wave. As a result, the characteristic curves are not straight lines (see Fig.~\ref{fig:characteristics}). We can think of Eq.~\eqref{GPEtoKPZ_2Burgers} as a pair of driven Burgers equations, in which the left moving waves act as a noise term on the propagation of the right moving waves, and vice versa. The viscous $\nu\partial_x^2v_{\pm}$ term is absent for this Hamiltonian system but will be generated upon coarse graining.

It is natural to ask how this situation changes for a \emph{Fermi} gas, which has the hydrodynamic description 
\begin{equation}
	\label{GPEtoKPZ_FermiHydro}
	H_{\text{Fermi}}=  \int dx \left[\frac{\rho\left(\partial_{x} \theta\right)^{2}}{2m}+\frac{\pi^{2}\rho^{3}}{6m}\right],
\end{equation}
with the second term representing the Fermi pressure. The same analysis now yields the \emph{uncoupled} Burgers equations
\begin{equation}
	\label{GPEtoKPZ_FermiBurgers}
	\partial_{t}v_{\pm}+v_{\pm}\partial_{x}v_{\pm} =0
\end{equation}
where $v_{\pm}=v\pm \pi \rho/m$ are the right and left moving Fermi velocities. The characteristics $X_{\pm}(t)$ are now straight lines, and the free Fermi gas therefore represents an exceptional fluid in which we expect no anomalous broadening of the type discussed here.


\emph{Numerical simulations}. The GPE is solved using the splitting method, whereby $\Psi(x,t)$ is evolved for a timestep $\tau$ alternately by the kinetic $T=\frac{1}{2}\int dx\, |\partial_{x}\Psi|^{2}$ and potential $V=\frac{1}{2}\int\ dx\,|\Psi|^{4}$ terms of the Hamiltonian \cite{McLachlan:1993}
\begin{equation}
	\label{GPEtoKPZ_TVsplitting}
\begin{split}
	&\cT_{\tau}:\tilde\Psi(k,t)\to e^{-ik^{2}\tau/2}\tilde\Psi(k,t)\\
	&\cV_{\tau}:\Psi(x,t)\to e^{-i\tau|\Psi(x,t)|^{2}}\Psi(x,t),
\end{split}
\end{equation}
where $\tilde\Psi(k,t)$ denotes the Fourier transform of $\Psi(x,t)$, and we now switch to measuring distance in units of the healing length $\xi$, time in units of (inverse) chemical potential $\mu\equiv g\rho_{0}$, and $\Psi$ in units of $\sqrt{\rho_{0}}$. Algorithms of this type are \emph{symplectic}. This means that the method exactly simulates a Hamiltonian $H_{\tau}$ with $H_{\tau}-H$ a power series in $\tau$. The lowest power of $\tau$ in the series determines the \emph{order} of the method. Two benefits of symplectic integrators for statistical mechanical simulations are: i) exact conservation of phase space volume (i.e. Liouville's theorem is satisfied) and ii) no drift in the energy due to exact conservation of $H_{\tau}$.

We use the method $\cV_{\tau/2}\cdot\cT_{\tau}\cdot\cV_{\tau/2}$ -- often called `Leapfrog' -- which is second order with \cite{McLachlan:1993}
\begin{equation*}
	\label{GPEtoKPZ_Htau}
	\begin{split}
	H_{\tau}-H &= \frac{\tau^{2}}{24}\left(2\{T,\{V,T\}\}+\{V,\{T,V\}\}\right)+O(\tau^{4})\\
	&=\frac{\tau^{2}}{24}\int dx\left[\rho^{2}\left(\partial_{x}\theta\partial_{x}^{3}\theta-(\partial_{x}^{2}\theta)^{2}\right)-\rho(\partial_{x}\rho)^{2}\right]\\
	&\qquad\qquad+O(\tau^{4})		
	\end{split}
\end{equation*}
%
We display the explicit form of the first correction term in $H_{\tau}$ to demonstrate that the additional terms generated by discretizing time are higher order in spatial gradients than Eq.~\eqref{GPEtoKPZ_CubicVertex}, and are therefore not expected to change the low $k$ behavior.


The harmonic modes are initially populated according to equipartition, so that the $\bop_{k}$ are taken as complex Gaussian random variables with $\langle|\bop_{k}|^{2}\rangle=\frac{T}{\cE_{k}}$ for temperature $T$. This assumes sufficiently weak nonlinearity, which is confirmed by the absence of transient behavior for the parameter ranges explored, indicating that the initial state is close to thermal.

\begin{figure}
	\centering
		\includegraphics[width=\columnwidth]{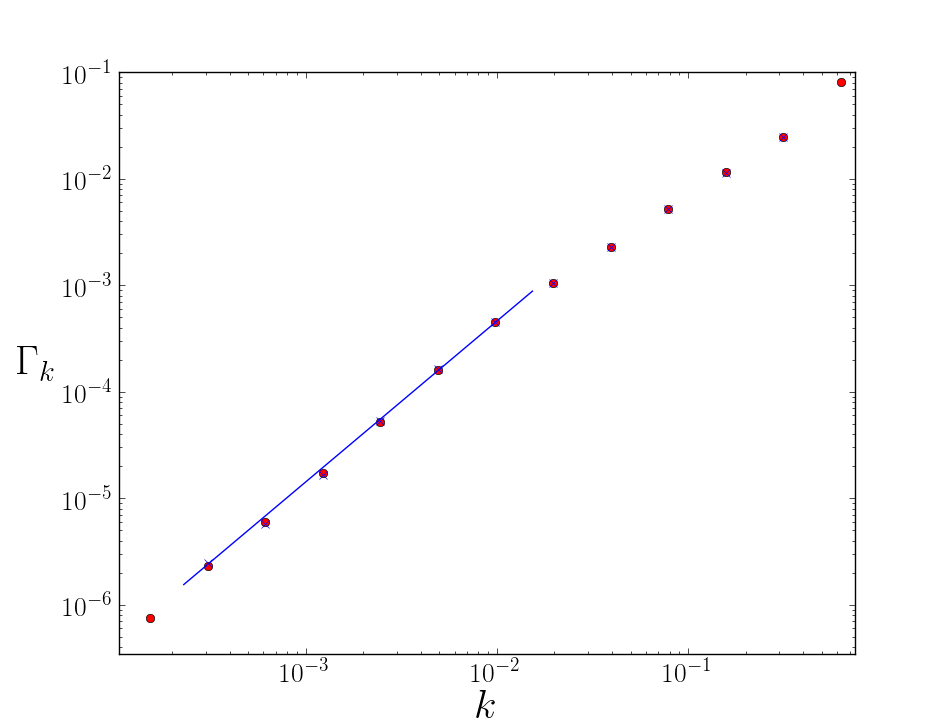}
	\caption{Dependence of linewidth on wavevector, for system sizes $N=2^{13}$ (red dots) and $N=2^{14}$ (blue crosses), with $T=0.005$. The data are almost perfectly coincident. 
	The blue line is a fit for wavevectors $2\pi p / L$, $p=2,4,\ldots 64$ giving $z = 1.510 \pm 0.018$. 
	Higher wavevectors show a marked deviation from $3/2$ scaling, and $k=2\pi/L$ is omitted because the linewidth lies below resolution.
	}
	\label{fig:widths}
\end{figure}

We choose a spatial discretization scale $a=5$ to ensure all wavevectors are in the regime of linear dispersion $k\ll \xi^{-1}$. The time step of $\tau=2$ is then at the limit of stability of the algorithm. Most simulations use systems of length $L=2^{13}a$, but we check that the results are not significantly altered for $L=2^{14}a$ (see Fig.~\ref{fig:widths}). Periodic boundary conditions are used throughout.

At each time step we compute the Fourier components of the density $\rho(x,t)$
\begin{equation}
	\label{GPEtoKPZ_rhoq}
	\rho_{k}(t) = \sum_{n=0}^{N-1} |\Psi(na,t)|^{2}e^{-2\pi i kna}\qquad k = 0,\frac{2\pi}{L},\ldots, \frac{\pi}{a}.
\end{equation}
The resulting time series is then Fourier transformed to give the dynamical structure factor
\begin{equation}
	\label{GPEtoKPZ_StructureFactor}
	S(k,\omega)=\langle |\rho_{k,\omega}|^{2}\rangle, \qquad \omega = 0,\pm\frac{2\pi}{N_{\text{bin}}\tau},\ldots
\end{equation}
where $N_{\text{bin}}$ is the bin size used for the computation of the power spectra (at least $2^{20}$), and the angular brackets denotes an average over  $\sim 128$ runs of length $N_{\text{steps}}\tau$ with different random initial conditions. Thus for $N_{\text{bin}}=2^{20}$,  $N_{\text{steps}}=2^{21}$  corresponds to an effective average over $\sim 256$ runs, assuming no correlation between the two halves of a given run. 

We gather data for wavevectors $k=\frac{2\pi p}{L}$, with $p=1,2, 4,\ldots 4096$. Typical results, displaying good data collapse (assuming a width scaling as $|k|^{3/2}$) are shown in Fig.~\ref{fig:Lineshape}. For an unbiased test, power spectra are folded assuming symmetry between positive and negative frequencies, and the phonon peaks fitted to a Lorentzian to extract the amplitude, peak frequency, and width. Data are shown in Fig.~\ref{fig:widths}. A good fit to the scaling form $\Gamma_{k}\propto |k|^{z}$ is obtained over 1.5 decades and yields $z = 1.510 \pm 0.018$. Significant deviations from scaling are obtained at higher wavevectors.



\emph{Conclusion}. We have provided convincing analytical and experimental evidence of the relationship between the finite temperature dynamics of the 1D Bose gas and the KPZ universality class. Galilean invariance is of paramount importance: simulations of wave equations with cubic nonlinearity but without Galilean invariance show $z\sim 1$ \cite{Kulkarni:}. Numerous extensions of the results of this work to multicomponent (or spinor) quantum fluids, and to transient rather than equilibrium dynamics, may be envisaged. In addition, the challenging problem of describing -- within a single framework -- the finite temperature phenomena described here, and the zero temperature results reviewed in Ref.~\cite{Imambekov:2011}, remains to be solved.


\emph{Acknowledgements} Our thanks are due to Robert McLachlan and Uwe T\"auber for helpful discussions, and to the University of Virginia Alliance for Computational Science and Engineering, especially Katherine Holcomb, for their assistance. A.L. gratefully acknowledges the support of the Research Corporation and the NSF through award DMR-0846788. M.K thanks Saul Lapidus for useful discussions.


%

\end{document}